\documentclass[a4paper,10pt,twoside]{cpc-hepnp}

\usepackage{multicol}
\usepackage{graphicx}
\usepackage{booktabs}
\usepackage{amssymb,bm,mathrsfs,bbm,amscd}
\usepackage[tbtags]{amsmath}
\usepackage{lastpage}

\newcommand{\be}{\begin{equation}}
\newcommand{\ee}{\end{equation}}
\newcommand{\bea}{\begin{eqnarray}}
\newcommand{\eea}{\end{eqnarray}}
\newcommand{\mev}{\mbox{MeV}}
\newcommand{\gev}{\mbox{GeV}}
\newcommand{\VVP}{\langle V V \! P\rangle}
\def\order#1{{\cal O}\left(#1\right)}

\newcommand{\power}[1]{\times 10^{#1}}
\newcommand{\ttcnew}[1]{\multicolumn{3}{c}{#1}}
\newcommand{\ttc}[1]{\multicolumn{2}{c}{#1}}

\begin{document}

\fancyhead[co]{\footnotesize A.~Nyf\/feler: Hadronic light-by-light scattering
contribution to the muon $g-2$} 

\footnotetext[0]{Received 31 December 2009}

\title{Hadronic light-by-light scattering contribution to the muon $g-2$}

\author{%
      A.~Nyf\/feler$^{1;1)}$\email{nyf\/feler@hri.res.in}%
}

\maketitle

\address{%
1~(Regional Centre for Accelerator-based Particle Physics, Harish-Chandra
Research Institute, \\ Chhatnag Road, Jhusi, Allahabad - 211019, India) \\ 
}

\begin{abstract}
We review recent developments concerning the hadronic light-by-light
scattering contribution to the anomalous magnetic moment of the muon. We first
discuss why fully off-shell hadronic form factors should be used for the
evaluation of this contribution to the $g-2$. We then reevaluate the
numerically dominant pion-exchange contribution in the framework of
large-$N_C$ QCD, using an off-shell pion-photon-photon form factor which
fulfills all QCD short-distance constraints, in particular, a new
short-distance constraint on the off-shell form factor at the external vertex
in $g-2$, which relates the form factor to the quark condensate magnetic
susceptibility in QCD. Combined with available evaluations of the other
contributions to hadronic light-by-light scattering this leads to the new
result $a_{\mu}^{\mathrm{LbyL; had}} = (116 \pm 40) \times 10^{-11}$, with a
conservative error estimate in view of the many still unsolved problems. Some
potential ways for further improvements are briefly discussed as well. For the
electron we obtain the new estimate $a_e^{\mathrm{LbyL; had}} = (3.9 \pm 1.3)
\times 10^{-14}$.
\end{abstract}

\begin{keyword}
muon, anomalous magnetic moment, hadronic contributions, effective field
theories, large-$N_C$ 
\end{keyword}

\begin{pacs}
14.60.Ef, 13.40.Em, 12.38.Lg
\end{pacs}

\begin{multicols}{2}

\section{Introduction}

The muon $g-2$ has served over many decades as an important test of the
Standard Model (SM). It is also sensitive to contributions from New Physics
slightly above the electroweak scale. For several years now a discrepancy of
about three standard deviations has existed between the SM prediction and the
experimental value, see the recent reviews Refs.~\cite{MdeRR07, FJ_Reviews,
JN09, new_had_VP} on the muon $g-2$. The main error in the theoretical SM
prediction comes from hadronic contributions, i.e.\ hadronic vacuum
polarization and hadronic light-by-light (had.\ LbyL) scattering. Whereas the
hadronic vacuum polarization contribution can be related to the cross section
$e^+ e^-\to~\mbox{hadrons}$, no direct experimental information is available
for had.\ LbyL scattering. One therefore has to rely on hadronic models to
describe the strongly interacting, nonperturbative dynamics at the relevant
scales from the muon mass up to about 2 GeV. This leads to large
uncertainties, see Refs.~\cite{BP07, PdeRV09, JN09} for recent reviews on
had.\ LbyL scattering.

The still valid picture of had.\ LbyL scattering as proposed some time back in
Ref.~\cite{deRafael94} is shown in Fig.~\ref{fig:had_LbyL}.
\begin{center}
\includegraphics[width=7cm]{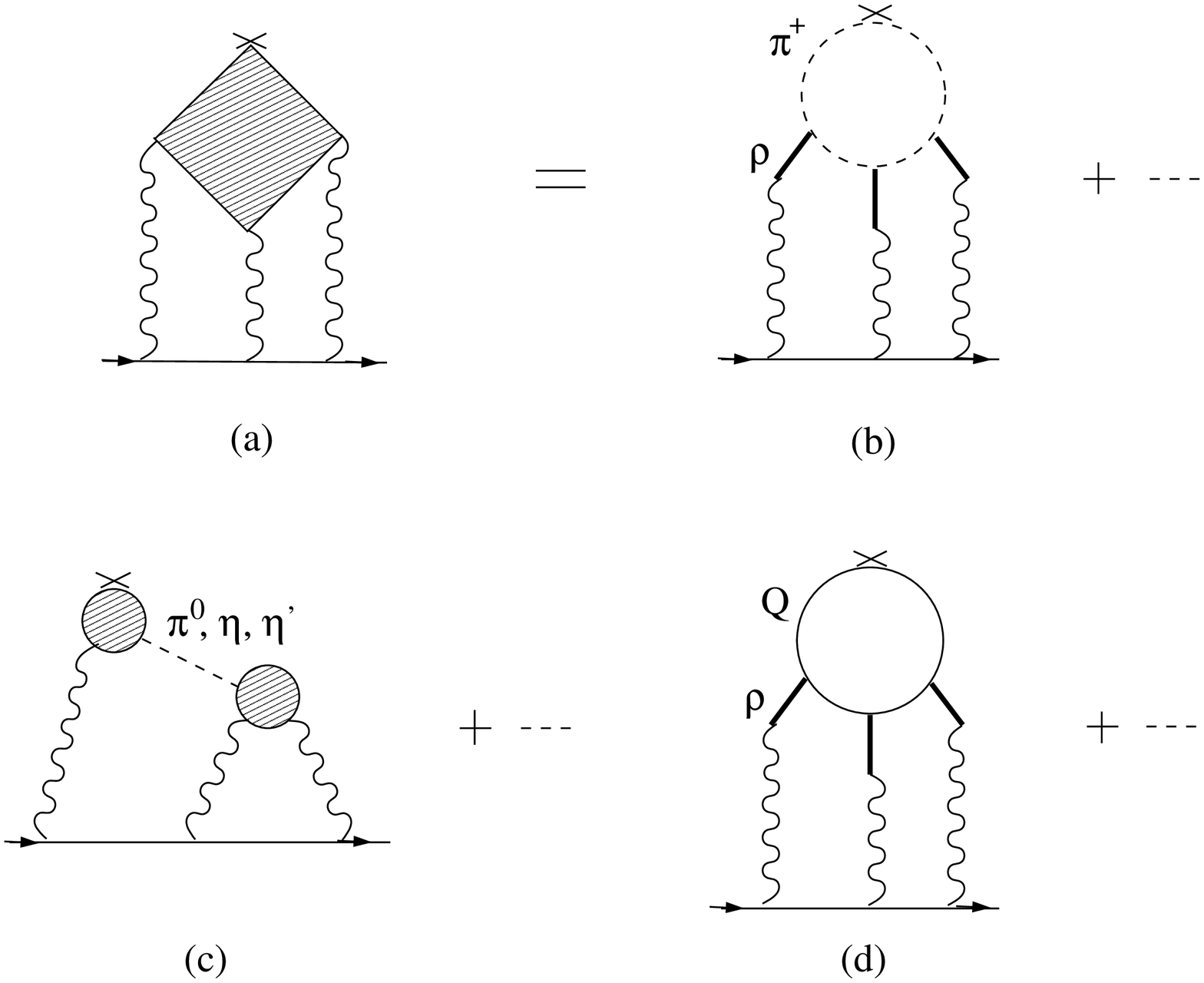}
\figcaption{The hadronic light-by-light scattering contribution to the muon
  $g-2$. 
\label{fig:had_LbyL}}
\end{center}
There are three classes of contributions to the relevant hadronic four-point
function $\langle VVVV\rangle$ [Fig.~\ref{fig:had_LbyL}(a)] which can also be
understood within an effective field theory approach to had.\ LbyL scattering:
(1) a charged pion and Kaon loop [Fig.~\ref{fig:had_LbyL}(b)], where the
coupling to photons is dressed by some form factor ($\rho$-meson exchange,
e.g.\ via vector meson dominance (VMD)), (2) pseudoscalar exchange diagrams
[Fig.~\ref{fig:had_LbyL}(c)] together with the exchanges of heavier resonances
($f_0, a_1, \ldots$) and, finally, (3) the irreducible part of the four-point
function which was modeled in Ref.~\cite{deRafael94} and later
works~\cite{BPP95, HKS95_HK98} by a constituent quark loop dressed with
VMD-type form factors [Fig.~\ref{fig:had_LbyL}(d)]. The latter contribution
can also be viewed as a short-distance complement of the employed low-energy
hadronic models. According to quark-hadron duality, the (constituent) quark
loop also models the contribution from the exchanges and loops of heavier
resonances, like $\pi^\prime, a_0^\prime, f_0^\prime, p, n, \ldots$, if they
are not explicitly included in the other terms.

One can try to reduce the model dependence and the corresponding uncertainties
by relating the hadronic form factors at low energies to results from chiral
perturbation theory and at high energies (short distances) to the operator
product expansion. In this way, one connects the form factors to the
underlying theory of QCD. This has been done in Refs.~\cite{BPP95, HKS95_HK98,
Bijnens_Persson01, KNetal01, MV03, Nyffeler09} for the numerically dominant
contribution from the exchange of light pseudoscalars $\pi^0, \eta$ and
$\eta^\prime$. In Ref.~\cite{MV03} also important short-distance constraints
on the axial-vector pole contribution have been imposed.

\section{On-shell versus off-shell form factors} 

It was pointed out recently in Ref.~\cite{FJ_Reviews}, that one should use
fully {\it off-shell} form factors for the evaluation of the LbyL scattering
contribution. This seems to have been overlooked in the recent literature, in
particular, in Refs.~\cite{Bijnens_Persson01,KNetal01,MV03,BP07,PdeRV09}.  The
on-shell form factors as used in Refs.~\cite{KNetal01,Bijnens_Persson01}
actually violate four-momentum conservation at the external vertex, as
observed already in Ref.~\cite{MV03}.

For illustration, we consider the contribution of the lightest intermediate
state, the neutral pion.  The key object which enters the diagram in
Fig.~\ref{fig:had_LbyL}(c) is the {\it off-shell} form factor ${\cal
F}_{{\pi^0}^*\gamma^*\gamma^*}((q_1+q_2)^2,q_1^2, q_2^2)$ which can be defined
via the QCD Green's function $\VVP$~\cite{BPP95,HKS95_HK98,Nyffeler09}
\bea
\lefteqn{ \hspace*{-0.5cm} \int d^4 x\, d^4 y \, e^{i (q_1 \cdot x + q_2 \cdot
    y)} \,  
\langle\,0 | T \{ j_\mu(x) j_\nu(y) P^3(0) \} | 0 \rangle \, } \nonumber \\  
&=& \Big[ \varepsilon_{\mu\nu\alpha\beta} \, q_1^\alpha q_2^\beta \, 
 {i \langle{\overline\psi}\psi\rangle \over F_\pi} \, {i \over (q_1 +
   q_2)^2 - m_\pi^2}  \nonumber \\
& & \ \times \ {\cal F}_{{\pi^0}^*\gamma^*\gamma^*}((q_1 +
 q_2)^2,q_1^2,q_2^2) \Big] \ + \ \ldots, 
\label{FFoffshellpi}
\eea
up to small mixing effects with the states $\eta$ and $\eta^\prime$ and
neglecting exchanges of heavier states like ${\pi^0}^\prime,
{\pi^0}^{\prime\prime}, \ldots$.  Here $j_\mu(x)$ is the light quark part of
the electromagnetic current and $P^3(x) = \left({\overline \psi} i \gamma_5
{\lambda^3 \over 2} \psi\right)(x)$. Note that for off-shell pions, instead of
$P^3(x)$, we could use any other suitable interpolating field, like
$\partial^\mu A_\mu^3(x)$ or even a fundamental pion field $\pi^3(x)$.

The corresponding contribution to the muon $g-2$ may be worked out with the
result~\cite{KNetal01}  
\end{multicols}
\ruleup
\vspace*{-0.2cm}
\bea
a_{\mu}^{\mathrm{LbyL};\pi^0} & = & - e^6
\int {d^4 q_1 \over (2\pi)^4} {d^4 q_2 \over (2\pi)^4}  
\,\frac{1}{q_1^2 q_2^2 (q_1 + q_2)^2[(p+ q_1)^2 - m_\mu^2][(p - q_2)^2 -
    m_\mu^2]} \nonumber \\
&& 
\qquad \times \left[
{{\cal F}_{{\pi^0}^*\gamma^*\gamma^*}(q_2^2, q_1^2, (q_1 + q_2)^2) \ {\cal 
    F}_{{\pi^0}^*\gamma^*\gamma}(q_2^2, q_2^2, 0) \over q_2^2 - 
m_{\pi}^2} \ T_1(q_1, q_2; p) \nonumber \right. \\
&& 
\qquad \quad  + \left. {{\cal F}_{{\pi^0}^*\gamma^*\gamma^*}((q_1+q_2)^2,
  q_1^2, q_2^2) \ {\cal F}_{{\pi^0}^*\gamma^*\gamma}((q_1+q_2)^2, (q_1+q_2)^2,
  0) \over (q_1+q_2)^2 - m_{\pi}^2} \ T_2(q_1, q_2; p) \right], 
\label{a_pion_2}
\eea
\ruledown \vspace{0.7cm}
\begin{multicols}{2}
\noindent
where the external photon has now zero four-momentum. See Ref.~\cite{KNetal01}
for the expressions for $T_i$.  Note that for general form factors a compact
three-dimensional integral representation for $a_{\mu}^{\mathrm{LbyL};\pi^0}$
has been derived in Ref.~\cite{JN09}.

Instead of the representation in Eq.~(\ref{a_pion_2}),
Refs.~\cite{Bijnens_Persson01,KNetal01} considered {\it on-shell} form factors
which would yield the so called {\it pion-pole} contribution, e.g.\ for the
term involving $T_2$, one would write 
\be
{\cal F}_{\pi^0 \gamma^* \gamma^*}(m_\pi^2,
q_1^2,q_2^2) \ \times \ {\cal F}_{\pi^0 \gamma^* \gamma}(m_\pi^2,
(q_1 + q_2)^2,0). 
\ee
Although pole dominance might be expected to give a reasonable approximation,
it is not correct as it was used in those references, as stressed in
Refs.~\cite{MV03,FJ_Reviews}.  The point is that the form factor
sitting at the external photon vertex in the pole approximation ${\cal
F}_{\pi^0 \gamma^* \gamma}(m_\pi^2,(q_1 + q_2)^2,0)$ for $(q_1 + q_2)^2 \neq
m_\pi^2$ violates four-momentum conservation, since the momentum of the
external (soft) photon vanishes.  The latter requires ${\cal F}_{{\pi^0}^*
\gamma^* \gamma}((q_1 + q_2)^2, (q_1 + q_2)^2, 0)$. Ref.~\cite{MV03} then
proposed to use instead 
\be
{\cal F}_{\pi^0 \gamma^* \gamma^*}(m_\pi^2,
q_1^2,q_2^2) \ \times \ {\cal F}_{\pi^0 \gamma \gamma}(m_\pi^2,
m_\pi^2,0)\,. 
\ee
Note that putting the pion on-shell at the external vertex automatically leads
to a constant form factor, given by the Wess-Zumino-Witten (WZW)
term~\cite{WZW}.  However, this prescription does not yield the {\it
  pion-exchange} contribution with off-shell form factors, which we calculate
with Eq.~(\ref{a_pion_2}).    

Strictly speaking, the identification of the pion-exchange contribution is
only possible, if the pion is on-shell. If one is off the mass shell of the
exchanged particle, it is not possible to separate different contributions to
the $g-2$, unless one uses some particular model where elementary pions can
propagate. In this sense, only the pion-pole contribution with on-shell form
factors can be defined, at least in principle, in a model-independent way.  On
the other hand, the pion-pole contribution is only a part of the full result,
since in general, e.g.\ using some resonance Lagrangian, the form factors will
enter the calculation with off-shell momenta.

\section{Pseudoscalar exchange contribution}

After the observation in Ref.~\cite{FJ_Reviews} that off-shell form factors
should be used, the numerically dominant pion-exchange contribution was
reanalyzed in detail in our paper~\cite{Nyffeler09}.  First we derived a new
QCD short-distance constraint on the off-shell pion-photon-photon form factor
${\cal F}_{{\pi^0}^* \gamma^* \gamma^*}$ from Eq.~(\ref{FFoffshellpi}) at the
external vertex in had.\ LbyL scattering. It arises in the limit when the
space-time argument of one of the vector currents in $\VVP$ approaches the
argument of the pseudoscalar density~\cite{KN_EPJC01}. 

In the chiral limit, assuming octet symmetry and up to corrections of order
$\alpha_s$, one then obtains the relation~\cite{Nyffeler09}
\be
\lim_{\lambda \to \infty} {\cal F}_{{\pi^0}^*\gamma^*\gamma}((\lambda
q_1)^2, (\lambda q_1)^2,0) = {F_0 \over 3} \ \chi 
+ \order{{1\over \lambda}}, 
\label{new_constraint}  
\ee
where $F_0$ is the pion-decay constant in the chiral limit and $\chi$ is the
  quark condensate magnetic susceptibility in QCD in the presence of a
  constant external electromagnetic field, introduced in
  Ref.~\cite{Ioffe_Smilga}: $\langle 0 | \bar{q} \sigma_{\mu\nu} q | 0
  \rangle_{F} = e \, e_q \, \chi \, \langle{\overline\psi}\psi\rangle_0 \,
  F_{\mu\nu}$, with $e_u = 2/3$ and $e_d = -1/3$.  Note that there is no
  falloff in Eq.~(\ref{new_constraint}) in this limit, unless
  $\chi$ vanishes.

Unfortunately there is no agreement in the literature what the actual value of
$\chi$ should be. Note that $\chi$ actually depends on the renormalization
scale $\mu$. Most recent estimates yield values $\chi(\mu = 1~\mbox{GeV})
\approx - 3~\mbox{GeV}^{-2}$~\cite{small_abs_chi}, although other approaches
give a much larger absolute value of $\chi(\mu = 0.5~\mbox{GeV}) \approx -
9~\mbox{GeV}^{-2}$~\cite{Ioffe_Smilga, large_abs_chi}.  While the running with
$\mu$ can explain part of the difference, it seems likely that the different
models used are not fully compatible.

In Ref.~\cite{Nyffeler09} we then reevaluated the pion-exchange contribution
using an off-shell form factor ${\cal F}_{{\pi^0}^*\gamma^*\gamma^*}((q_1 +
q_2)^2, q_1^2, q_2^2)$ in the framework of large-$N_C$ QCD.  In the spirit of
the minimal hadronic Ansatz~\cite{MHA} for Green's functions in large-$N_C$
QCD, such a form factor had already been constructed in
Ref.~\cite{KN_EPJC01}. It generalizes the usual VMD form factor and contains
the two lightest multiplets of vector resonances, the $\rho$ and the $\rho'$
(lowest meson dominance (LMD) +V). In contrast to the VMD ansatz, the LMD+V
form factor fulfills all the relevant short-distance constraints derived in
Refs.~\cite{KN_EPJC01, MV03, Nyffeler09}, including the new one from
Eq.~(\ref{new_constraint}). In Ref.~\cite{Nyffeler09} we assumed that the
LMD/LMD+V framework is self-consistent, therefore the estimate $\chi^{\rm LMD}
= - 2 / M_V^2 = -3.3~\mbox{GeV}^{-2}$ was used (with a typical large-$N_C$
uncertainty of 30\%), which is compatible with other
estimates~\cite{small_abs_chi}.

Other model parameters are fixed by normalizing the form factor to the pion
decay amplitude $\pi^0 \to \gamma\gamma$ and by reproducing experimental
data~\cite{CELLO90_CLEO98} for the on-shell form factor ${\cal
F}_{\pi^0\gamma^*\gamma}(m_\pi^2, -Q^2, 0)$, see Ref.~\cite{Nyffeler09} for
all the details.  Recently, BABAR~\cite{BABAR09} has published new data for
this form factor which does not show the characteristic falloff for large
Euclidean momentum, $\lim_{Q^2 \to \infty} \: {\cal F}_{\pi^0 \gamma^*
\gamma}(m_\pi^2,-Q^2,0) \sim 1 / Q^2$~\cite{LepageBrodsky80}. Some
implications of this new experimental result have been been discussed in
Ref.~\cite{Chernyak09}. As shown in Ref.~\cite{Nyffeler_Chiral_Dynamics},
using the BABAR data to fit some of the LMD+V model parameters, does, however,
not change the final result given below.

Varying all the LMD+V model parameters in reasonable ranges and adding all
uncertainties linearly to cover the full range of values obtained with the
scan of parameters, we get the new estimate~\cite{Nyffeler09}
\be \label{amupi0LMD+V}
a_\mu^{\mathrm{LbyL};\pi^0} = (72 \pm 12) \times 10^{-11}.
\ee

As far as the contribution to $a_\mu$ from the exchanges of the other light
pseudoscalars $\eta$ and $\eta^\prime$ is concerned, a simplified approach was
adopted in Ref.~\cite{Nyffeler09}, as was done earlier in other
works~\cite{BPP95,HKS95_HK98,KNetal01,MV03}. We took a simple VMD form factor,
normalized to the experimental decay width $\Gamma(\mbox{PS} \to \gamma
\gamma)$. In this way one obtains the results $a_{\mu}^{\mathrm{LbyL};\eta} =
14.5 \times 10^{-11}$ and $a_{\mu}^{\mathrm{LbyL};\eta^\prime} = 12.5 \times
10^{-11}$. Adding up the contributions from all the light pseudoscalar
exchanges, we obtain the new estimate~\cite{Nyffeler09}
\be \label{amuLbLPS}
a_{\mu}^{\mathrm{LbyL;PS}} = (99 \pm 16) \times 10^{-11}, 
\ee
where we have assumed a 16\% error, as inferred above for the pion-exchange
contribution. 

For comparison, we have listed in Table~\ref{tab:pseudoscalars} some
evaluations of the pion- and pseudoscalar-exchange contribution to had.\ LbyL
scattering by various groups. The model used by each group has also been
indicated in the first column of the table, see the corresponding references
for all the details (we used the abbreviations: ENJL = Extended
Nambu-Jona-Lasinio model; HLS = Hidden Local Symmetry model; $\chi$QM = chiral
quark model; FF = form factor; $h_2$ is one of the LMD+V model parameters).

\end{multicols}

\begin{center}
\tabcaption{ \label{tab:pseudoscalars} Results for the $\pi^0, \eta$ and 
  $\eta^\prime$ exchange contributions obtained by various groups.}
\footnotesize
\begin{tabular}{lr@{(}c@{)}lr@{(}c@{)}l}
\hline
 Model for ${\cal F}_{P^{(*)}\gamma^*\gamma^*}$
 &\multicolumn{3}{c}{$a_\mu^{\mathrm{LbyL}; \pi^0}\power{11}$} & 
\multicolumn{3}{c}{$a_\mu^{\mathrm{LbyL; PS}}\power{11}$}\\
\hline
Point coupling & \ttcnew{$\hspace*{-0.35cm}+\infty$} &
\ttcnew{$\hspace*{-0.3cm}+\infty$} \\ 
modified ENJL (off-shell) [BPP]~\cite{BPP95} & \hspace*{0.5cm}59& 9 &
& \hspace*{0.5cm}85&13& \\ 
VMD / HLS (off-shell) [HKS, HK]~\cite{HKS95_HK98}  & 57&4& & 83&6& \\
nonlocal $\chi$QM (off-shell) [DB]~\cite{Dorokhov_Broniowski} & 65 & 2 & &
\ttcnew{\hspace*{-0.3cm}$-$} \\  
AdS/QCD (off-shell ?) [HoK]~\cite{AdSQCD} & \ttcnew{\hspace*{-0.72cm}$69$} & 
\ttcnew{\hspace*{-0.83cm}$107$} \\
LMD+V (on-shell, $h_2=0$) [KN]~\cite{KNetal01} & 58&10& & 83&12& \\
LMD+V (on-shell, $h_2=-10~\gev^2$) [KN]~\cite{KNetal01} & 63&10& &
88&12& \\ 
LMD+V (on-shell, constant FF at external vertex) [MV]~\cite{MV03} 
& 77 & 7 & & 114 & 10 \\  
LMD+V (off-shell) [N]~\cite{Nyffeler09}   & 72 & 12 & & 99 & 16 \\  
\hline
\end{tabular}
\end{center}

\begin{multicols}{2}

Our results for the pion and the sum of all pseudoscalar exchanges are about
20\% larger than the values in Refs.~\cite{BPP95, HKS95_HK98} which used other
hadronic models that presumably do not obey the new short-distance constraint
from Eq.~(\ref{new_constraint}) and thus have a stronger damping at large
momentum. Within the non-local $\chi$QM used in
Ref.~\cite{Dorokhov_Broniowski} there is a strong, exponential suppression for
large pion virtualities. According to Ref.~\cite{AdSQCD}, the estimate with
the AdS/QCD model has an error of at most 30\%. On the other hand, our result
is smaller than the pion- and pseudoscalar-pole contribution calculated in
Ref.~\cite{MV03}.  Since only the {\it pion-pole} contribution is considered
in Ref.~\cite{MV03}, their short-distance constraint cannot be directly
applied to our approach. However, our ansatz for the pion-exchange
contribution agrees qualitatively with the short-distance behavior of the
quark-loop derived in Ref.~\cite{MV03}, see the discussion in
Refs.~\cite{Nyffeler09, JN09}.  Note, however, that the numerical value for
the pion-pole contribution listed as [MV] in Table~\ref{tab:pseudoscalars}
should rather be $80 \times 10^{-11}$, see Refs.~\cite{BP07,
Dorokhov_Broniowski, Nyffeler09}.

\section{Summary of other contributions}

In Table~\ref{tab:summary} we have collected the results for all the
contributions to had.\ LbyL scattering according to Fig.~\ref{fig:had_LbyL}
obtained by various groups in recent times, including some ``guesstimates''
for the total value. In the following, we highlight the main features of the
numbers given and point out some critical issues regarding each
contribution. A more detailed discussion can be found in Ref.~\cite{JN09}.

\end{multicols}

\ruleup 

\vspace*{-0.3cm}
\begin{center}
\tabcaption{Summary of the most recent results for the various contributions
  to $a_{\mu}^{\mathrm{LbyL; had}} \times 10^{11}$.  The last column is our
  estimate based on our new evaluation for the pseudoscalars and some of the
  other results.
\label{tab:summary}} 
\footnotesize
\begin{tabular}{cr@{$\pm$}lr@{$\pm$}lr@{$\pm$}lr@{$\pm$}lr@{$\pm$}lr@{$\pm$}lr@{$\pm$}l}
\hline
Contribution & \ttc{\hspace*{0.2cm}BPP~\cite{BPP95}} &
\ttc{\hspace*{-0.1cm}HKS, HK~\cite{HKS95_HK98}} & 
\ttc{\hspace*{-0.1cm}KN~\cite{KNetal01}} & \ttc{MV~\cite{MV03}} &
\ttc{BP~\cite{BP07}, MdRR~\cite{MdeRR07}} & \ttc{PdRV~\cite{PdeRV09}} &
\ttc{\hspace*{-0.1cm}N~\cite{Nyffeler09}, JN~\cite{JN09}} \\  
\hline
 $\pi^0,\eta,\eta'$ & $85$ & $13$ & $82.7 $&$ 6.4$ & $ 83 $&$ 12$
& $114 $&$ 10$ & \ttc{$-$} & $114$ & $13$ & $99$&$16$ \\
axial vectors & $2.5 $&$ 1.0$ & $1.7 $&$ 1.7$ &\ttc{\hspace*{-0.15cm}$-$} & 
$22 $&$ 5$& \ttc{$-$} & $15$ & $10$ & $22 $&$ 5$  \\ 
scalars  & $-6.8$&$ 2.0$ & \ttc{\hspace*{-0.1cm}$-$} &
 \ttc{\hspace*{-0.15cm}$-$} &  \ttc{$-$}& \ttc{$-$} &  $-7$ & $7$ & $-7$&$2$
 \\  
$\pi,K$ loops & $-19$ &$ 13$ &  $-4.5$&$ 8.1$ &\ttc{\hspace*{-0.13cm}$-$} &
 \ttc{$-$} &  \ttc{$-$} & $-19$ & $19$ &$-19 $&$ 13$ \\ 
$\pi,K~\mbox{loops} \atop + \mbox{subl.}~N_C$ & \ttc{\hspace*{0.2cm}$-$} &
 \ttc{\hspace*{-0.15cm}$-$} & 
\ttc{\hspace*{-0.15cm}$-$} & $0$&$ 10$ & \ttc{$-$} &
 \ttc{\hspace*{0.1cm}$-$} & \ttc{\hspace*{-0.35cm}$-$} \\  
quark loops  & $21 $&$ 3$ & $9.7 $&$ 11.1$ & \ttc{\hspace*{-0.13cm}$-$} &
\ttc{$-$} & \ttc{$-$} &  \ttc{\hspace*{-0.05cm}$2.3$} & $21 $&$ 3$\\
\hline 
Total & $83$&$ 32$ & $89.6 $&$ 15.4$ & $80 $& $40$ &
$136$ & $25$ & \hspace*{0.6cm}$110$ & $40$ & $105$ & $26$ & $116$ & $39$  \\ 
\hline
\end{tabular}

\end{center}

\vspace*{0.5cm}
\ruledown \vspace*{0.2cm} 
\begin{multicols}{2}

As one can see from Table~\ref{tab:summary}, the different models used by
various groups lead to slightly different results for the individual
contributions.  The final result is dominated by the pseudoscalar exchange
contribution, which is leading in large-$N_C$, but subleading in the chiral
counting. The other contributions are smaller, but not
negligible. Furthermore, they cancel out to some extent, in particular the
dressed pion and Kaon loops and the dressed quark loops.

In Ref.~\cite{MV03}, new QCD short-distance constraints were derived for the
axial-vector {\it pole} contribution with on-shell form factors ${\cal
F}_{A\gamma^*\gamma^*}$ at both vertices. A huge enhancement of a factor of
ten was observed compared to the earlier estimates in Refs.~\cite{BPP95,
HKS95_HK98} which assumed nonet symmetry for the states $a_1, f_1$ and
$f_1^\prime$. It was shown that the result is very sensitive to the mass of
the exchanged axial-vector resonance. Since the form factors include light
vector mesons like the $\rho$, this leads to a smaller effective mass of the
exchanged resonance, compared to $M_A \sim 1300~\mev$.  The result is also
sensitive to the mixing of the states $f_1$ and $f_1^\prime$. The result given
in Table~\ref{tab:summary} corresponds to ideal mixing. If $f_1$ is a pure
octet state and $f_1^\prime$ a pure singlet, the final result goes down to
$a_\mu^{\mathrm{LbyL}; a_1, f_1, f_1^\prime} = 17 \times 10^{-11}$.  The
procedure adopted in Ref.~\cite{MV03} is an important improvement over
Refs.~\cite{BPP95, HKS95_HK98} and we have therefore taken the result for the
axial-vectors from that reference for our final estimate for the full had.\
LbyL scattering contribution. This despite the fact that only on-shell form
factors have been used in Ref.~\cite{MV03}. As we argued above, we think that
one should use consistently off-shell form factors at the internal and the
external vertex.

Within the ENJL model used in Ref.~\cite{BPP95}, the scalar exchange
contribution is related via Ward identities to the constituent quark loop. In
fact, Ref.~\cite{HKS95_HK98} argued that the effect of the exchange of scalar
resonances below several hundred MeV might already be included in the sum of
the (dressed) quark loops and the (dressed) pion and Kaon loops. Such a
potential double-counting is definitely an issue for the broad sigma meson
$f_0(600)$. It is also not clear which scalar resonances are described by the
ENJL model used in Ref.~\cite{BPP95}. The parameters were determined from a
fit to various low-energy observables and resonance parameters, among them a
scalar multiplet with mass $M_S = 983~\mbox{MeV}$. However, with those fitted
parameters, the ENJL model actually predicts a rather low mass of $M_S^{\rm
ENJL} = 620~\mbox{MeV}$. 

The (dressed) charged pion- and Kaon-loops from Fig.~\ref{fig:had_LbyL}(b)
yield the leading contribution in the chiral counting, but are subleading in
$N_C$. We note that the result without dressing (scalar QED) is actually
finite: $a_\mu^{\mathrm{LbyL}; \pi^\pm} = - 46 \times 10^{-11}$.  The dressing
with form factors then leads to a rather huge and very model dependent
suppression (compare the results for Refs.~\cite{BPP95} and \cite{HKS95_HK98}
in Table~\ref{tab:summary}), so that the final result is much smaller than the
one obtained for the pseudoscalars. This effect was studied in
Ref.~\cite{MV03} for the HLS model used in Ref.~\cite{HKS95_HK98}, in an
expansion in $(m_\pi / M_\rho)^2$. Ref.~\cite{MV03} observed a large
cancellation between the first few terms in the series and the expansion
converges only very slowly. The main reason is that typical momenta in the
loop integral are of order $\mu = 4 m_\pi \approx 550~\mbox{MeV}$ and the
effective expansion parameter is $\mu / M_\rho$. The authors of
Ref.~\cite{MV03} took this as an indication that the final result is very
likely suppressed, but also very model dependent and that the chiral expansion
looses its predictive power. The pion and Kaon loops contribution is then only
one among many potential contributions of ${\cal O}(1)$ in $N_C$ and they lump
all of these into the guesstimate $a_\mu^{\mathrm{LbyL}; N_C^0} = (0 \pm 10)
\times 10^{-11}$. However, since this estimate does not even cover the results
for the pion and Kaon loops given in Refs.~\cite{BPP95,HKS95_HK98}, we think
this procedure is not very appropriate.

The (dressed) constituent quark loops from Fig.~\ref{fig:had_LbyL}(d) are also
leading in large-$N_C$. The result with point-like couplings is finite:
$a_\mu^{\mathrm{LbyL; quarks}} = 62 \times 10^{-11}$. The dressing with form
factors then leads again to a large and very model dependent suppression of
the final result, compare Refs.~\cite{BPP95} and \cite{HKS95_HK98} in
Table~\ref{tab:summary}.

In the recent review~\cite{PdeRV09} the central values of some of the
individual contributions to had.\ LbyL scattering were adjusted and some
errors were enlarged to cover the results obtained by various groups which
used different models, see Table~\ref{tab:summary}.  Finally, the errors were
added in quadrature. Maybe the resulting small error masks some of the
uncertainties we still face in had.\ LbyL scattering. Note that the dressed
light quark loops are not included as a separate contribution in
Ref.~\cite{PdeRV09} (only the contribution from a bare $c$-quark is included
in Table~\ref{tab:summary}). The light quark loops are assumed to be already
covered by using the short-distance constraint from Ref.~\cite{MV03} on the
pseudoscalar-{\it pole} contribution. Although numerically the final estimate
from Ref.~\cite{MV03} is very close to our result given in
Table~\ref{tab:summary}, in view of the interpretation given for this term in
the Introduction, we do not see any reason, why the contribution from the
dressed quark loops should be discarded completely. At least in large-$N_C$
QCD, only the sum of {\it all} resonance exchanges should be dual to the quark
loops.

\section{Conclusions}

Combining our result for the pseudoscalars with the evaluation of the
axial-vector contribution in Ref.~\cite{MV03} and the results from
Ref.~\cite{BPP95} for the other contributions, we obtain the new 
estimate~\cite{Nyffeler09, JN09}  
\be
a_{\mu}^{\mathrm{LbyL; had}} = (116 \pm 40) \times 10^{-11} 
\ee
for the total had.\ LbyL scattering contribution to the anomalous magnetic
moment of the muon.\footnote{Applying the same procedure to the electron, we
get $a_e^{\mathrm{LbyL};\pi^0} = (2.98 \pm 0.34) \times
10^{-14}$~\cite{Nyffeler09}. Note that the naive rescaling
$a_e^{\mathrm{LbyL};\pi^0}(\mathrm{rescaled}) = (m_e / m_\mu)^2 \
a_\mu^{\mathrm{LbyL};\pi^0} = 1.7 \times 10^{-14}$ yields a value which is
almost a factor of 2 too small. Our estimates for the other pseudoscalars
contributions are $a_e^{\mathrm{LbyL};\eta} = 0.49 \times 10^{-14}$ and
$a_e^{\mathrm{LbyL};\eta^\prime} = 0.39 \times 10^{-14}$. Therefore we get
$a_e^{\mathrm{LbyL;PS}} = (3.9 \pm 0.5) \times 10^{-14}$. Assuming that the
pseudoscalar contribution yields the bulk of the result of the total had.\
LbyL scattering correction, we obtain $a_e^{\mathrm{LbyL;had}} = (3.9 \pm 1.3)
\times 10^{-14}$, with a conservative error of about 30\%, see
Ref.~\cite{JN09}. This value was later confirmed in the published version of
Ref.~\cite{PdeRV09} where a leading logs estimate yielded
$a_e^{\mathrm{LbyL;had}} = (3.5 \pm 1.0) \times 10^{-14}$.}  The variation of
the results for the individual contributions listed in Table~\ref{tab:summary}
reflects our inherent ignorance of strong interaction physics in had.\ LbyL
scattering. One can take the differences between those values as an indication
of the model uncertainty and, to be conservative, all the errors have been
added linearly, as was done earlier in Refs.~\cite{BPP95, KNetal01, BP07,
MdeRR07}.

Certainly, more work on the had.\ LbyL scattering contribution is needed to
fully control all the uncertainties, in particular, if we want to fully profit
from a potential future $g-2$ experiment with an expected error of about $15
\times 10^{-11}$~\cite{future_g-2_exp}. Maybe at some point we will get an
estimate from lattice QCD~\cite{LbyL_lattice_QCD}, although the relevant QCD
Green's function $\langle VVVV \rangle$, to be integrated over the phase space
of three off-shell photons, is a very complicated object.

In the meantime we suggest the following way forward~\cite{JN09}. It is very
important to have a unified framework (hadronic model) which deals with {\it
all} the contributions to had.\ LbyL scattering. A purely phenomenological
approach would be to use some resonance Lagrangian where all couplings are
fixed from experiment. Since such Lagrangians are in general
non-renormalizable it is, however, not clear how to achieve a proper matching
with QCD at short distances. Such a matching can be achieved within the
large-$N_C$ framework, however, the corresponding resonance Lagrangians in
general contain many unknown coefficients and it will be difficult to fix all
of them theoretically or experimentally. In any case, in both of these
approaches any additional experimental information on various hadronic form
factors would be very useful to constrain the theoretical models. In this
respect, $e^+ e^-$ colliders running at energies around $0.5 - 2~\gev$ could
help to measure some of the form factors relevant for had.\ LbyL
scattering~\cite{FF_from_exp}.

\vspace*{0.2cm} 

\acknowledgments{I would like to thank the organizers of PHIPSI09 for the
  invitation and their financial support. I am grateful to F.\ Jegerlehner for
  many helpful discussions and numerous correspondences. This work was
  partially supported by funding from the Department of Atomic Energy,
  Government of India, for the Regional Centre for Accelerator-based Particle
  Physics (RECAPP), Harish-Chandra Research Institute.}

%
%

\end{multicols}



\vspace{0.3cm}

\begin{thebibliography}{90}

\vspace*{0.15cm} 

\bibitem{MdeRR07}
J.~P.~Miller, E.~de Rafael, and B.~L.~Roberts,
Rep.\ Prog.\ Phys.\ {\bf 70}, 795 (2007). 


\bibitem{FJ_Reviews}
F.~Jegerlehner,
Acta Phys.\ Pol.\  B {\bf 38}, 3021 (2007); 
%
F.~Jegerlehner, \textit{The Anomalous Magnetic Moment of the Muon},
Springer Tracts Mod.\ Phys.\ Vol.\ 226 (Springer, Berlin, 2008). 


\bibitem{JN09}
  F.~Jegerlehner and A.~Nyffeler,
  Phys.\ Rept.\  {\bf 477}, 1 (2009). 


\bibitem{new_had_VP}
For recent updates of the hadronic vacuum polarization contribution, see: 
  M.~Davier {\it et al.}, 
  arXiv:0908.4300 [hep-ph]; 
%
M.~Davier, talk at this conference; T.~Teubner, talk at this conference. 


\bibitem{BP07}
J.~Bijnens and J.~Prades,
Mod.\ Phys.\ Lett.\ A {\bf 22}, 767 (2007). 


\bibitem{PdeRV09}
J.~Prades, E.~de Rafael, and A.~Vainshtein in {\it Lepton Dipole Moments},
B.L.\ Roberts and W.J.\ Marciano, (eds) (World Scientific, Singapore, 2009),
309, arXiv:0901.0306 [hep-ph]. 


\bibitem{deRafael94}
  E.~de Rafael,
  Phys.\ Lett.\  B {\bf 322}, 239 (1994). 


\bibitem{BPP95}
J.\ Bijnens, E.\ Pallante, and J.\ Prades, 
Phys.\ Rev.\ Lett.\ {\bf 75}, 1447 (1995); {\bf 75}, 3781(E) (1995);  
%
%
Nucl.\ Phys.\ {\bf B474}, 379 (1996); 
%
{\bf B626}, 410 (2002).


\bibitem{HKS95_HK98} 
M.~Hayakawa, T.~Kinoshita, and A.~I.~Sanda,
Phys.\ Rev.\ Lett.\  {\bf 75}, 790 (1995); 
%
%
Phys.\ Rev.\ D {\bf 54}, 3137 (1996); 
%
M.\ Hayakawa and T.\ Kinoshita, 
Phys.\ Rev.\ D {\bf 57}, 465 (1998); {\bf 66}, 019902(E) 
  (2002). 


\bibitem{KNetal01}
M.~Knecht and A.~Nyffeler,
Phys.\ Rev.\ D {\bf 65}, 073034 (2002);  
%
M.~Knecht {\it et al.}, 
Phys.\ Rev.\ Lett.\  {\bf 88}, 071802 (2002). 


\bibitem{MV03}
K.~Melnikov and A.~Vainshtein,
Phys.\ Rev.\ D {\bf 70}, 113006 (2004). 


\bibitem{Bijnens_Persson01}
  J.~Bijnens and F.~Persson,
  hep-ph/0106130.


\bibitem{Nyffeler09}
  A.~Nyffeler,
  Phys.\ Rev.\  D {\bf 79}, 073012 (2009). 


\bibitem{WZW}
J.\ Wess and B.\ Zumino, Phys.\ Lett.\  {\bf 37B}, 95 (1971); 
E.\ Witten, Nucl.\ Phys.\ {\bf B223}, 422 (1983). 


\bibitem{KN_EPJC01}
  M.~Knecht and A.~Nyffeler,
  Eur.\ Phys.\ J.\  C {\bf 21}, 659 (2001). 


\bibitem{Ioffe_Smilga} 
  B.~L.~Ioffe and A.~V.~Smilga,
  Nucl.\ Phys.\  {\bf B232}, 109 (1984).


\bibitem{small_abs_chi}
  I.~I.~Balitsky and A.~V.~Yung,
  Phys.\ Lett.\  {\bf 129B}, 328 (1983); 
%
  I.~I.~Balitsky, A.~V.~Kolesnichenko, and A.~V.~Yung,
  Yad.\ Fiz.\  {\bf 41}, 282 (1985); 
%
  V.~Y.~Petrov {\it et al.}, 
  Phys.\ Rev.\  D {\bf 59}, 114018 (1999); 
%
  P.~Ball, V.~M.~Braun, and N.~Kivel,
  Nucl.\ Phys.\  {\bf B649}, 263 (2003);  
  A.~E.~Dorokhov,
  Eur.\ Phys.\ J.\  C {\bf 42}, 309 (2005); 
%
  V.~Mateu and J.~Portoles,
  Eur.\ Phys.\ J.\  C {\bf 52}, 325 (2007); 
%
  J.~Rohrwild,
  JHEP {\bf 0709}, 073 (2007); 
%
  K.~Goeke {\it et al.}, 
  Phys.\ Rev.\  D {\bf 76}, 116007 (2007); 
%
  B.~L.~Ioffe,
  Phys.\ Lett.\  B {\bf 678}, 512 (2009); 
%
  P.~V.~Buividovich {\it et al.}, 
  Nucl.\ Phys.\  B {\bf 826}, 313 (2010). 


\bibitem{large_abs_chi}
  V.~M.~Belyaev and Y.~I.~Kogan,
  Yad.\ Fiz.\  {\bf 40}, 1035 (1984); 
%
  A.~Vainshtein,
  Phys.\ Lett.\  B {\bf 569}, 187 (2003); 
%
  S.~Narison,
  Phys.\ Lett.\  B {\bf 666}, 455 (2008). 


\bibitem{MHA}
  B.~Moussallam and J.~Stern,
  hep-ph/9404353;
%
  B.~Moussallam,
  Phys.\ Rev.\  D {\bf 51}, 4939 (1995);  
%
  B.~Moussallam,
  Nucl.\ Phys.\  {\bf B504}, 381 (1997);  
%
S.~Peris, M.~Perrottet, and E.~de Rafael,
J.\ High Energy Phys.\ 05 (1998) 011; 
%
M.~Knecht {\it et al.}, 
Phys.\ Rev.\ Lett.\  {\bf 83},  5230 (1999). 


\bibitem{CELLO90_CLEO98}
H.~J.~Behrend {\it et al.}  [The CELLO Collaboration],
Z.\ Phys.\ C {\bf 49}, 401 (1991); 
%
J.~Gronberg {\it et al.}  [The CLEO Collaboration],
Phys.\ Rev.\ D {\bf 57}, 33 (1998). 


\bibitem{BABAR09}
  B.~Aubert {\it et al.}  [The BABAR Collaboration],
  Phys.\ Rev.\  D {\bf 80}, 052002 (2009); 
%
V.~P.~Druzhinin, talk at this conference. 


\bibitem{LepageBrodsky80}
G.~P.~Lepage and S.~J.~Brodsky,
Phys.\ Rev.\ D {\bf 22},  2157 (1980);
S.~J.~Brodsky and G.~P.~Lepage, 
Phys.\ Rev.\ D {\bf 24}, 1808 (1981). 


\bibitem{Chernyak09}
  V.~L.~Chernyak,
  arXiv:0912.0623 [hep-ph], these proceedings. 


\bibitem{Nyffeler_Chiral_Dynamics}
  A.~Nyffeler,
  arXiv:0912.1441 [hep-ph].


\bibitem{Dorokhov_Broniowski}
  A.~E.~Dorokhov and W.~Broniowski,
  Phys.\ Rev.\  D {\bf 78}, 073011 (2008). 


\bibitem{AdSQCD} 
  D.~K.~Hong and D.~Kim,
  Phys.\ Lett.\  B {\bf 680}, 480 (2009). 


\bibitem{future_g-2_exp}
B.~L.~Roberts, Nucl.\ Phys.\ B (Proc.\ Suppl.) {\bf 131}, 157 (2004); 
%
R.~M.~Carey {\it et al}, Proposal of the BNL Experiment E969, 2004; 
J-PARC Letter of Intent L17; 
%
D.~W.~Hertzog, Nucl.\ Phys.\ Proc.\ Suppl.\  {\bf 181-182}, 5 (2008); 
%
B.~L.~Roberts, talk at this conference; T.~Mibe, talk at this conference. 


\bibitem{LbyL_lattice_QCD}
  M.~Hayakawa {\it et al.}, 
  PoS {\bf LAT2005}, 353 (2006)  [arXiv:hep-lat/0509016]; 
%
  T.~Blum and S.~Chowdhury,
  Nucl.\ Phys.\ Proc.\ Suppl.\  {\bf 189}, 251 (2009).


\bibitem{FF_from_exp}
G.~Venanzoni, talk at this conference. 

\end{thebibliography}
\end{document}